# A comparative study of the D0 neural-network analysis of the top quark non-leptonic decay channel

R.Odorico

University of Bologna, Department of Physics and Istituto Nazionale di Fisica Nucleare, Sezione di Bologna
Via Irnerio 46, I-40126 Bologna, Italy
e-mail: odorico@www.bo.infn.it

**Abstract**. A simpler neural-network approach is presented for the analysis of the top quark non-leptonic decay channel in events of the D0 Collaboration. Results for the top quark signal are comparable to those found by the D0 Collaboration by a more elaborate handling of the event information used as input to the neural network.

## 1 Introduction

The D0 Collaboration has recently presented an analysis of its Fermilab Tevatron collider data, aimed to establish a $t\bar{t}$ signal with both top quarks decaying non-leptonically. The analysis makes use of neural networks (NN), whose input is mainly represented by thoroughly researched variables built from a complete and detailed jet reconstruction for the events [1,2].

This analysis, like previous theoretical NN studies of the same signal (see, e.g., [3]), is based on the presumption that NN's should be "helped" by insight we gather from the theoretical event representations developed to interpret the events. In other words, it is left aside the possibility that the NN acting directly on the raw data may find internal representations by itself, which are adequate for discrimination.

The study presented here consists of a NN application with the latter approach: instead of using jets as intermediaries, the NN is directly applied to a calorimetric deposition grid. Simulated events for $t\bar{t}$ and the QCD background are provided by the event generator COJETS 6.24 [4]. The NN is a conventional feed-forward network with back-propagation training [5] handled by program NEURAL 2.0 [6].

By mimicking the event selection criteria used by D0, it is found that the NN results for $t\bar{t} \to$ non-leptonic with this direct and simpler approach are comparable to those obtained by D0 employing jet-reconstruction.

## 2 Event selection

$t\bar{t}$ and QCD background events are provided by the event generator COJETS 6.24 [4]. For $p\bar{p}$ interactions at a c.m. energy $E_{cm}$ = 1.8 TeV and a top quark mass $m_t$ = 175 GeV it yields a total $t\bar{t}$ cross-section $\sigma(t\bar{t})$ = 7.31 pb.

A series of cuts is imposed in order to: i) suppress the contribution of leptonic and semi-leptonic decays of the top quark; ii) reduce the amount of the background. Where possible, the cuts mimic those used in the D0 analysis.

For the suppression of the top leptonic and semi-leptonic decay contributions, events containing an electron or muon with a transverse momentum $p_T$ > 20 GeV or a missing transverse energy $E_T^{miss}$ > 20 GeV are excluded.

For the sake of background reduction the following requirements are imposed:

i) Total transverse energy $E_T^{tot}$ > 120 GeV;
ii) At least one muon with transverse momentum $p_T$ > 4 GeV and absolute pseudorapidity $|\eta|$ < 2.5.
iii) A linear aplanarity $A_L$ > 0.5. Considering only particles with $|\eta|$ < 2.5 (neutrinos excluded), after finding the thrust, or major axis $\vec{M}$, in the transverse energy plane, and defining the minor axis $\vec{m}$ as the axis orthogonal to it, linear aplanarity is defined as $A_L = \sum_i |\vec{E}_i \cdot \vec{m}| / \sum_i |\vec{E}_i \cdot \vec{M}|$, where $\vec{E}_i$ is the transverse energy vector of particle $i$.

For the integrated luminosity $\mathcal{L}$ = 110.2 pb$^{-1}$ applying to the D0 event sample, the events left after all the cuts are:
i) 128 events for $t\bar{t}$; ii) 3868 events for the background.

## 3 Variables

The region $|\eta|$ < 2.5 is covered by a grid of 40 calorimetric cells with $\Delta\eta$ x $\Delta\varphi$ = 1 x 45°. Their energy depositions are thresholded at 10 GeV, i.e. cells with a deposition less than that are set to deposition zero. That is in order to suppress cumulative effects from the many cells with low depositions, which may disorient the NN. The origin of the azimuth $\varphi$ is set, for each event, by the side of the minor axis $\vec{m}$ opposite to that of $\sum_i \vec{E}_i \cdot \vec{m}$.

Thus, the variables we use as input to the NN are:

i) The 40 variables given by the energy depositions in the calorimetric cells;
ii) The total (visible) transverse energy $E_T^{tot}$;
iii) The maximum muon transverse momentum observed in the event, $p_{T,\max}^{\mu}$.

## 4 Neural network

The neural network NN is structured in 2 cascading feed-forward nets, both with 1 hidden layer and 1 output node. The first, NN$_1$, has in input only the 40 calorimetric deposition variables and employs 8 hidden nodes. The NN$_1$ output, $E_T^{tot}$ and $p_{T,\max}^{\mu}$ are the input variables of the second feed-forward net, NN$_2$, which employs 2 hidden nodes. Its output, which in the following will be referred to as the NN output, varies between 0, the background target, and 1, the top signal target.

The back-propagation training is applied first to NN$_1$, with outputs from the trained net applied to the training events stored away. The same set of events (5,000 for each type) is then used for the back-propagation training of NN$_2$.

## 5 Results

For a comparison with the NN results of the D0 analysis, we pick up the dependence on the NN output of the ratio $R = N(t\bar{t} \to \text{non-leptonic}) / N(\text{Background})$, where numerator and denominator represent the numbers of events specified in parentheses. Fig. 1 shows $R$ versus the NN output from the D0 results and from this study.

The results are comparable taking into account that, besides the intrinsic uncertainties of the Monte Carlo simulations involved in both calculations, in this study no special effort has been made in optimizing the event cuts, the calorimetric segmentation (pretty rough, although economical, as it stands now), the choice of the variables, the NN structure. That would have been outside the scope of this study, meant to theoretically examine the viability of this alternative approach and not aimed to a thorough quantitative interpretation of experimental data, like in the case of the D0 analysis.

## 5 Conclusions

Instead of employing an elaborate jet reconstruction, in a neural network analysis of the events it may be convenient to directly use the calorimetric map: the neural network itself will care to build its own internal representations of the data to discriminate.

It should be noted that in the place of the deliberately down-to-earth handling of the calorimetric map used here, one can extract variables from it by using tools from the well supplied mathematical arsenal for the analysis of 2-dimensional data (like, e.g., the currently popular wavelet transform).

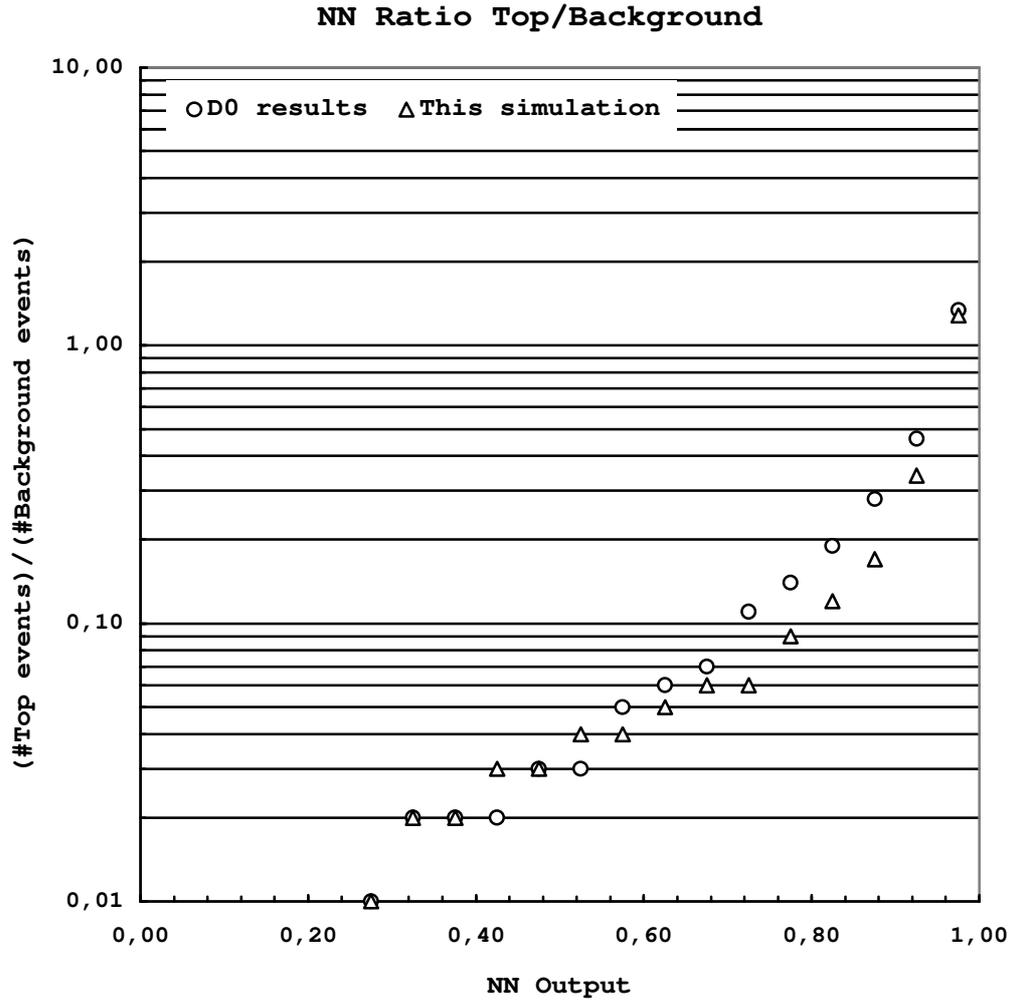

**Fig 1**. Dependence of the ratio $R = N\,(t\bar{t} \to$ non-leptonic$) / N$ (Background) on the NN output, according to the results of the D0 analysis and of this study.